\documentstyle[12pt,fleqn,epsf]{article}
\renewcommand{\baselinestretch}{1.2}

\def\fnote#1#2{\begingroup\def\thefootnote{#1}\footnote{#2}\endgroup}
\makeatletter
\def\section{\@startsection {section}{1}{\z@}{3.5ex plus 1ex minus
    .2ex}{2.3ex plus .2ex}{\sc }}
\def\subsection{\@startsection{subsection}{2}{\z@}{3.25ex plus 1ex
minus
   .2ex}{1.5ex plus .2ex}{\small \sc }}
\def\appendix{\par\clearpage
  \setcounter{section}{0}
  \setcounter{subsection}{0}
  \@addtoreset{equation}{section}
  \def\@sectname{Appendix~}
  \def\theequation{\thesection.\arabic{equation}}
  \def\thesection{\Alph{section}}}
\makeatother
\makeatletter \@addtoreset{equation}{section} \makeatother
\renewcommand{\theequation}{\thesection.\arabic{equation}}

\def\ap#1#2#3{     {\it Ann. Phys. (NY) }{\bf #1} (19#2) #3}

\def\npb#1#2#3{    {\it Nucl. Phys. }{\bf B #1} (19#2) #3}

\def\plb#1#2#3{    {\it Phys. Lett. }{\bf B #1} (19#2) #3}
\def\prd#1#2#3{    {\it Phys. Rev. }{\bf D #1} (19#2) #3}

\def\prl#1#2#3{    {\it Phys. Rev. Lett. }{\bf #1} (19#2) #3}
\def\ptp#1#2#3{    {\it Prog. Theor. Phys. }{\bf #1} (19#2) #3}

\def\zpc#1#2#3{    {\it Z. Physik }{\bf C #1} (19#2) #3}

\def\nc#1#2#3{     {\it Nuovo Cim. }{\bf #1} (19#2) #3}

\def\ibid#1#2#3{   {\it ibid. }{\bf #1} (19#2) #3}

\def\eq#1{{eq.~(\ref{#1})}}
\def\eqs#1#2{{eqs.~(\ref{#1})--(\ref{#2})}}

\let\vev\VEV

\def\Im{\mathop{\mbox{Im}}}
\def\Re{\mathop{\mbox{Re}}}
\def\Tr{\mathop{\mbox{Tr}}\,}

\def\GG{$\vev{\alpha_s GG/ \pi}$ }
\newcommand{\bea}{\begin{eqnarray}}
\newcommand{\beq}{\begin{equation}}
\newcommand{\eea}{\end{eqnarray}}
\newcommand{\eeq}{\end{equation}}
\newcommand{\nnu}{\nonumber}
\newcommand{\spav}[1]{\parbox{1mm}{\vspace*{#1}}}

\begin{document}
\begin{titlepage}

\spav{0.3cm}
\begin{center}
{\Large\bf The Physics of $K^0$-$\bar K^0$ Mixing:}\\ 
{\large\bf $\widehat{B}_{K}$ and $\Delta M_{L S}$ in the Chiral Quark Model}\\
\spav{0.5cm}\\
{\large V. Antonelli, S. Bertolini, 
M. Fabbrichesi and
E.I. Lashin\fnote{\dag}{Permanent address:
Ain Shams University, Faculty of Science, Dept. of Physics, Cairo, Egypt.}}
\spav{1.0cm}\\
{\em  INFN, Sezione di Trieste}\\
{\em and}\\ 
{\em Scuola Internazionale Superiore di Studi Avanzati}\\
{\em via Beirut 4, I-34013 Trieste, Italy.}\\
\end{center}

\noindent
\hrulefill

{\small 
\noindent 
{\sc \bf Abstract}\\[-.3cm]

\noindent
We compute the  $\widehat B_{K}$ parameter and the mass difference 
$\Delta M_{LS}$ of the $K^0$-$\bar K^0$ system by means of 
the chiral quark model.
The chiral coefficients of the relevant $\Delta S=2$ and $\Delta S=1$ chiral
lagrangians are computed via quark-loop integration. 
We include the relevant effects  
of one-loop corrections in chiral perturbation theory.
The final result is very sensitive to non-factorizable corrections
of $O(\alpha_S N)$ coming from gluon condensation.
The size of the gluon condensate is determined
by fitting the experimental value of the amplitude 
$K^+ \rightarrow \pi^+\pi^0$.
By varying all the relevant parameters we obtain 
$$\widehat{B}_{K}= 0.87\pm 0.33\ .$$
We evaluate within the model the long-distance contributions to 
$\Delta M_{LS}$ induced by the double insertion of the  $\Delta S = 1$ 
chiral lagrangian and study the interplay between short- and long-distance
amplitudes. By varying all parameters we obtain 
$$\Delta M_{LS}^{th}/\Delta M_{LS}^{exp} = 0.76\ ^{+0.64}_{-0.34}\ .$$
Finally,
we investigate the phenomenological constraints on the Kobayashi-Maskawa
parameter Im $\lambda _t$ entering the determination of
$\varepsilon'/\varepsilon$.}

\noindent
\hrulefill

\vfill

\flushleft{\tt SISSA 20/96/EP \\[-0.28ex]
September 1996}

%

\end{titlepage}

\newpage
\setcounter{footnote}{0}
\setcounter{page}{1}

The mixing in the $K^0$-$\bar K^0$ system 
is determined by the weak effective hamiltonian
through the mass matrix
$\langle \bar K^0 | {\cal H}_W | K^0 \rangle$ .  Its computation
requires  both
 long- and short-distance physics. For this reason,
 the $CP$ violating quantity $\varepsilon$ and 
 the mass difference $\Delta M_{LS}$,
which are respectively related to the
imaginary and the real part of the mass matrix, are
factorized in the renormalization-group invariant
parameter $\widehat{B}_K$, which takes into account
long-distance effects, and the Wilson coefficients, which account for
the short distance ones.
$\Delta M_{LS}$ receives in addition
a long-distance correction from the double insertion of the
$\Delta S=1$ hamiltonian.

In this work we apply 
the chiral quark model~\cite{QM}($\chi$QM) and chiral perturbation
techniques to estimate the long-distance part of the  mass matrix.
Such an analysis follows 
our recent studies of the
$\Delta I = 1/2$ selection rule~\cite{II} and 
$\varepsilon'/\varepsilon$~\cite{III} within
the $\chi$QM approach to kaon physics.

The long-distance contributions
thus computed are eventually
 matched to an up-to-date next-to-leading order (NLO) 
determination of the short-distance coefficients and the results
 compared with the experimental values.

\section {Introduction}	

The effective $\Delta S=2$ quark 
lagrangian at scales $\mu<m_c$ is given by
\bea
{\cal L}_{\Delta S = 2} &=& -
\frac{G_F^2 M_{W}^2 }{4 \pi^2} \left[\lambda_c^2 \eta_{1} S(x_c) 
+ \lambda_t^2 \eta_{2} S(x_t)
 + 2 \lambda_c \lambda_t \eta_3 S(x_c , x_t)\right] b(\mu) Q_{S2} (\mu)
\ , \nnu \\
& &
\label{lags2}
\eea
where $G_F$ is the Fermi constant, $M_W$ is the $W$ boson mass,
$x_i = m_i^2 / M_W^2$, and $\mu$ is the renormalization scale.
The parameters 
$
\lambda_j = V_{jd} V_{js}^*
$
represent the relevant combinations of Kobayashy-Maskawa (KM) 
matrix elements ($j=u,c,t$). Finally we denote by
$Q_{S2}$ the $\Delta S=2$ local four quark operator
\beq
Q_{S2} =(\bar{s}_L \gamma^{\mu} d_L) (\bar{s}_L \gamma_{\mu} d_L)
\, .
\eeq

The integration of the electroweak loops leads to the
 Inami-Lim functions \cite{Inami-Lim}
\beq
S(x) = x \left[ \frac{1}{4} + \frac{9}{4} \frac{1}{1-x} 
-\frac{3}{2} \frac{1}{(1-x)^2} \right] -\frac{3}{2} 
\left[ \frac{x}{1- x} \right] ^3 \ln x 
\, ,
\label{inimf1}
\eeq
\beq
S(x_c, x_t) =  -x_c \ln x_c + x_c \left[
\frac{x_t^2 -8 x_t +4}{4 (1 - x_t)^2} \ln x_t 
+ \frac{3}{4} \frac{x_t}{x_t - 1} \right]
\, 
\label{inimf2}
\eeq
which depend on the masses of the charm and top quarks and describe the 
$\Delta S = 2$ transition amplitude in the absence of strong interactions.

The short-distance QCD corrections are encoded in the coefficients $\eta_1$,
$\eta_2$ and $\eta_3$ with a common scale-dependent
factor $b(\mu)$ factorized out.
They are functions of the heavy quarks masses and of the scale parameter
 $\Lambda_{\rm QCD}$.
These QCD corrections are available to NLO~\cite{BJW,HH1,HH2} in the strong and
electromagnetic couplings.

The scale-dependent common factor of the short-distance corrections 
is given by
\beq
b ( \mu ) = \left[\alpha_s\left(\mu\right)\right]^{-2/9}
\left( 1 - J_3 \frac{\alpha_s\left(\mu\right)}{4 \pi} \right)
\, ,
\label{wc}
\eeq
where $J_3$ depends on the $\gamma_5$-scheme used in the regularization. The
naive dimensional regularization (NDR) gives
\beq
J_3^{\rm NDR} = -\frac{307}{162} \, ,
\eeq
 while in the 't Hooft-Veltman (HV) scheme one finds
 \beq
 J_3^{\rm HV} =-\frac{91}{162} \, .
 \eeq
 
For the running QCD coupling we take the average over 
recent LEP and SLC determinations~\cite{alfa}, 
\beq
\alpha_S(m_Z) = 0.119 \pm 0.006
\label{alfar}
\eeq 
which corresponds to
\beq   
\Lambda_{\rm QCD}^{(4)}= 350 \pm 100 \: \mbox{MeV} \, .
\label{lambdone}
\eeq 

The scale-dependent $B_K (\mu)$ parameter is defined by the matrix element 
\beq
\langle \bar{K^0}|Q_{S2}\left(\mu\right)|K^0 \rangle = \frac{4}{3} f_K^2 m_K^2
B_K (\mu) 
\, ,
\label{bk}
\eeq              
where $f_K$ and $m_K$ are the kaon decay constant and mass, respectively
(see table 2 for their numerical values).

The value of $B_K (\mu)$ measures the deviation of the matrix
element from the  vacuum
saturation approximation used in the original work of 
Gaillard and Lee~\cite{GL}, 
namely $B_K(\mu)=1$.
The physically relevant parameter is
 $\hat{B}_K$, which is defined by the relation:
\beq
\hat{B}_K = B_K(\mu) b(\mu) \, .
\eeq
This quantity should be in principle renormalization
scale independent.
As we include the perturbative NLO determination of the Wilson
coefficient, we shall also
discuss  the $\gamma_5$-scheme dependence of our
result. 

\begin{table}
\begin{center}
\begin{tabular}{|c|c|c|}
\hline $\widehat{B}_K$ & Method & Reference\\ 
\hline
$3/4$ & Leading $1/N_c$ & \cite{BG86} \\
$0.37$ & Lowest-Order Chiral Perturbation Theory & \protect\cite{DGH}\\ 
\hline
$0.70\pm0.10$ & Next-to-Leading $1/N_c$ Estimate &\protect\cite{BBG}\\
$0.4\pm0.2$ & Next-to-Leading $1/N_c$ Estimate, $O(p^2)$ &
\protect\cite{PdeR}\\
$0.42\pm 0.06$ & $O(p^4)$ Chiral Perturbation Theory &
\protect\cite{Bruno}\\
$0.60\div 0.80$ & NJL model with spin-1 interactions &
\protect\cite{Bijnens-Prades}
\\ \hline
$0.39\pm0.10$ & QCD-Hadronic Duality &
\protect\cite{PR85,PDPPR91}\\
$0.5\pm 0.1\pm0.2$ & QCD Sum Rules (3-Point Functions) &
\protect\cite{BDG88}\\
$0.55\pm 0.25$ & QCD Sum Rules (3-Point Functions) &
\protect\cite{Decker}\\
$0.58\pm 0.22$ & Laplace Sum Rule &
\protect\cite{Narison}
\\ \hline
$ 0.90\pm 0.03\pm 0.14$ & Lattice
& \protect\cite{SH96} \\ \hline
\end{tabular}
\end{center}
\caption{Values of $\widehat{B}_K$ obtained in different approaches.}
\label{tabbk}
\end{table}

A useful up-to-date summary of various determinations of
this parameter is given in Table 1 which updates that of ref.~\cite{PP}.

We have followed   the approach
described in ref.~\cite{HME} in which
the weak chiral lagrangian is considered
as the effective theory of  the $\chi$QM~\cite{QM}.
In the present case, it is the bosonization of the operator  $Q_{S2}$
and the determination of the coefficient of the corresponding
$\Delta S=2$ chiral lagrangian that is made possible by the $\chi$QM.

In the determination of $B_K(\mu)$ to $O(\alpha_s N_c)$
enters the contribution of the gluon condensate.
The final estimate is very sensitive to the value of
such an input parameter. In order to restrict the range of 
allowed values, we
impose the additional constraint of taking for the gluon condensate
the value that gives the best fit of the experimental
 amplitude  $K^+ \rightarrow \pi^+\pi^0$, which is related at the
leading order in chiral perturbation theory to that of
$K^0 \rightarrow \bar K^0$. Such a procedure is  consistent
with that followed in ref.~\cite{II} where we reproduced the
$\Delta I = 1/2$ rule by a similar choice of input parameters.      

A ``long-distance'' scale dependence is introduced by the  
one-loop chiral corrections to the hadronic matrix elements. 
According to our approach, this scale dependence
should match that in the Wilson coefficients and provide a 
scale independent value of $\widehat B_K$ within the uncertainties
quoted for higher order corrections. 
In the case of $\widehat B_K$, we find that there cannot be matching at this
order insofar as both the Wilson coefficient and the chiral corrections
renormalize the parameter in the same direction. The scale dependence
remains however below 20\% and the final estimate is thus still reliable.

Our approach is in principle
sensitive to the scheme used to treat $\gamma_5$ matrices in a generic
space-time dimension.
The NDR prescription of ref.~\cite{Roma,Monaco}
 preserves the chiral properties of the 
operator $Q_{S2}$ by means of 
a convenient normalization of  the evanescent operators. 
As discussed in ref.~\cite{HME}, the consistency with such a prescription makes
the matrix elements of $Q_{S2}$ the same
in the two schemes. 
As a consequence, the remnant scheme dependence of the final result
is that present in the short-distance factor $b(\mu)$.
 
There are two important parameters related to the $K^{0}$-$\bar{K}^{0}$ 
mixing: the $CP$ violating quantity $\varepsilon$ which is proportional to the
imaginary part of the mass matrix and the mass difference 
$\Delta M_{LS} \equiv m_L - m_S$.
 The observed value for these 
quantities are ~\cite{PDG}:
\beq
 |\varepsilon| =  (2.266 \pm 0.023) \times 10^{-3} \label{EPS}  
\eeq
and
\beq
\Delta M_{LS} = (3.510 \pm 0.018) \times 10^{-15}\ \mbox{GeV} \, . 
\label{MD}
\eeq
Knowing $\varepsilon$, we can determine $\Im \lambda_t$, as discussed in
section 6.
As a by-product of the computation one also obtains an estimate for
the width difference $\Delta \Gamma_{LS}$, the experimental value of which is
\beq
\Delta \Gamma_{LS} = -(7.374  \pm 0.010) \times 10^{-15}\ \mbox{GeV}\, . 
\label{GAMD}
\eeq
However, a consistent determination of this quantity requires one 
extra order in
perturbation theory, as we shall discuss below.

From the theoretical point of view,  
the $K^0-\bar K^0$ mass matrix can be written, using CPT invariance, as
\bea
{\cal M}& = & \frac{1}{2 m_K} \, \, 
\left( \begin{array}{cc} \langle K^{0} |H_W|K^{0}\rangle & 
\langle K^{0} |H_W|\bar{K}^{0}\rangle \\
\langle \bar{K}^{0} |H_W|K^{0}\rangle & 
\langle \bar{K}^{0} |H_W|\bar{K}^{0}\rangle                             
\end{array} \nnu
\right) \\
 & = &
\left( \begin{array}{cc} M - \frac{1}{2} i \Gamma & 
M_{12} -\frac{1}{2} i \Gamma_{12} \\
M^*_{12} - \frac{1}{2} i \Gamma^*_{12} & 
M - \frac{1}{2} i \Gamma                             
\end{array}
\right)
\label{mmatrix}
\eea

In the presence of $CP$ violation ($\varepsilon\neq 0$) $M_{12}$ and
$\Gamma_{12}$ are complex numbers.
The diagonalization of the  mass matrix (~\ref{mmatrix}) 
leads to the physical states:
\beq
\begin{array}{ccc}
K_{L} &=& \frac{1}{\sqrt{2 (1+|\varepsilon|^2)}} 
          \left[ K^{0} (1+\varepsilon) + \bar{K}^{0} (1-\varepsilon)\right] \\
K_{S} &=& \frac{1}{\sqrt{2 (1+|\varepsilon|^2)}} 
          \left[ K^{0} (1+\varepsilon) - \bar{K}^{0} (1-\varepsilon)\right]
\end{array}
\label{pstate}
\eeq
For a tiny $CP$ violation,
their associated mass and width differences are given by:
\beq
\Delta M_{LS} = 
 \frac{1}{m_{K}} \Re\,\left[\langle K^{0} |H_W|\bar{K}^{0}\rangle\right] 
\label{MDform}
\ ,
\eeq

\beq
\Delta \Gamma_{LS}= 
- \,\, \frac{2}{m_K} \Im\, \left[\langle K^{0} |H_W|\bar{K}^{0}\rangle\right] 
\label{GD}
\,.
\eeq

In order to estimate these two parameters we need to evaluate
in addition to  the
quark box-diagram contribution, coming from the $\Delta S =2$  
effective weak lagrangian given in (\ref{lags2}), 
the long-distance contribution coming from the double insertion of
the $\Delta S=1$ weak chiral lagrangian. 
In the latter case,
the mixing between $K^{0}$ and $\bar{K}^{0}$ 
can proceed,
up to the one-loop level, 
via one- and two-particle intermediate states
\beq
K^{0} \rightarrow \left(\pi^{0},\eta\right)\to \bar{K}^{0},
\label{1pi}
\eeq
\beq
K^{0} \rightarrow \left(\pi^{+} \pi^{-},K^{+} K^{-},\pi^{0}\pi^{0},\eta\eta,
\pi^{0}\eta\right)\to \bar{K}^{0},
\label{tpi}
\eeq
 
Within the $\chi$QM approach the $\Delta S=1$ weak chiral lagrangian
can be systematically derived at a given order in momentum expansion
starting from the  effective quark lagrangian~\cite{GW1}:
\beq
{\cal L}_{\Delta S = 1} =
- \frac{G_F}{\sqrt{2}} V_{ud} V_{us}^{*}  \sum_i \Bigl[
z_i(\mu) + \tau y_i(\mu) \Bigr] Q_i (\mu)
 \, , \label{ham}
\eeq
where $Q_i$ are local four-quark operators obtained by 
integrating out in the standard
model the vector bosons and the heavy quarks $t,\,b$ and $c$. A convenient
and by now standard basis includes the following ten quark operators:
 \beq
\begin{array}{rcl}
Q_{1} & = & \left( \overline{s}_{\alpha} u_{\beta}  \right)_{\rm V-A}
            \left( \overline{u}_{\beta}  d_{\alpha} \right)_{\rm V-A}
\, , \\[1ex]
Q_{2} & = & \left( \overline{s} u \right)_{\rm V-A}
            \left( \overline{u} d \right)_{\rm V-A}
\, , \\[1ex]
Q_{3,5} & = & \left( \overline{s} d \right)_{\rm V-A}
   \sum_{q} \left( \overline{q} q \right)_{\rm V\mp A}
\, , \\[1ex]
Q_{4,6} & = & \left( \overline{s}_{\alpha} d_{\beta}  \right)_{\rm V-A}
   \sum_{q} ( \overline{q}_{\beta}  q_{\alpha} )_{\rm V\mp A}
\, , \\[1ex]
Q_{7,9} & = & \frac{3}{2} \left( \overline{s} d \right)_{\rm V-A}
         \sum_{q} \hat{e}_q \left( \overline{q} q \right)_{\rm V\pm A}
\, , \\[1ex]
Q_{8,10} & = & \frac{3}{2} \left( \overline{s}_{\alpha} 
                                                 d_{\beta} \right)_{\rm V-A}
     \sum_{q} \hat{e}_q ( \overline{q}_{\beta}  q_{\alpha})_{\rm V\pm A}
\, , 
\end{array}  
\label{Q1-10} 
\eeq
where  $\alpha$, $\beta$ denote color indices ($\alpha,\beta
=1,\ldots,N_c$) and $\hat{e}_q$  are quark charges. Color
indices for the color singlet operators are omitted. 
$(V\pm A)$ refer to
$\gamma_{\mu} (1 \pm \gamma_5)$.
We recall that
$Q_{1,2}$ stand for the $W$-induced current--current
operators, $Q_{3-6}$ for the
QCD penguin operators and $Q_{7-10}$ for the electroweak penguin (and box)
ones. 

The functions $z_i(\mu)$ and $y_i(\mu)$ are the
 Wilson coefficients and $V_{ij}$ the KM matrix elements; $\tau = - V_{td}
V_{ts}^{*}/V_{ud} V_{us}^{*}$. 

In a previous work~\cite{HME} we have computed the chiral coefficients
for the complete $O(p^2)$ $\Delta S=1$ chiral lagrangian. We will make use
of those results to evaluate the long-distance contributions in
\eqs{1pi}{tpi}.

\section{A model independent estimate of $\widehat B_K$}

The $\Delta S = 2$ matrix element can be related via chiral symmetry
to that of the $\Delta S =1$ and $\Delta I = 3/2$ amplitude
${\cal A}(K^+\to\pi^+\pi^0)$\cite{DGH}.
Neglecting the $SU(3)$ breaking effects related to
the chiral loop corrections to the matrix element,
the electromagnetic contributions and the $\pi-\eta$ mixing, 
we obtain the relation
\beq
\frac{4}{3} f_K^2 m_K^2 \widehat B_K =
\frac{\sqrt{2}}{G_F}\frac{f_\pi}{V_{us}^*V_{ud}}\frac{m_K^2}{m_K^2-m_\pi^2}
\frac{b(\mu)}{z_1(\mu)+z_2(\mu)}\ {\cal A}(K^+\to\pi^+\pi^0)\ .
\label{bkdi}
\label{mie}
\eeq
In the previous equation $V_{us}$ and $V_{ud}$ are two matrix elements of 
the KM
mixing matrix, $b(\mu)$ is the $\Delta S=2$ Wilson coefficient given by 
(\ref{wc}), while $z_1(\mu)$ and $z_2(\mu)$ are the real parts of the Wilson 
coefficients for the two $\Delta S=1$ operators $Q_1$ and $Q_2$ which 
dominate the $K^+\to\pi^+\pi^0$ transition.

By inputting the experimental value 
${\cal A}(K^+\to\pi^+\pi^0) = 1.84\times 10^{-8}$ GeV
and the NLO results for the Wilson coefficients (the ratio
$b(\mu)/(z_1(\mu)+z_2(\mu))$ is to a large extent $\mu$ and $\gamma_5$-scheme
independent) we find the model ``independent'' estimate
\beq
\widehat B_K = 0.40\ .
\label{BKmi}
\eeq
This number updates the value $\widehat B_K = 0.33$
given in ref. \cite{DGH}. 

On the other hand, having a model that reproduces the experimental result, 
in order to apply correctly \eq{bkdi} we must subtract in 
${\cal A}(K^+\to\pi^+\pi^0)$ all the chiral symmetry breaking
corrections due to chiral loops, 
electroweak penguins and $\pi-\eta$ mixing~\cite{II}.
In this way we obtain in the $\chi$QM approach, on the basis of chiral 
symmetry arguments alone, the following $O(p^2)$ prediction: 
\beq
\widehat B_K =\frac{3}{4} b\left(\mu\right) \frac{f_\pi f}{f_K^2} 
\left[1+\frac{1}{N_c} \left(1-\delta_{\langle GG \rangle} \right)\right] \, . 
\label{BK0}
\eeq
In the previous formula we have denoted by $\delta_{\langle GG \rangle}$
the non-perturbative gluonic corrections which arise in the 
$\chi$QM approach,  
\beq
\delta_{\langle GG \rangle} = \frac{N_c}{2} \frac{\langle
\alpha_s G G/\pi \rangle}{16 \pi^2 f^4} \, ,
\label{gg}
\eeq
where \GG is the gluon condensate
and $N_c$ is the number of colors.
We will come back to these corrections in the next 
section. 

In considering \eq{BK0} it is important to remember that 
the factor $f_ \pi$ comes  from the soft pion theorem, while $f$ is 
the chiral lagrangian parameter appearing in the  
calculation of the amplitude ${\cal A}(K^+\to\pi^+\pi^0)$.
At the tree level $f=f_\pi$.
The spurious $\mu$ dependence present in \eq{BK0} should be canceled
by that of the hadronic matrix elements, which is absent at the
lowest order in the chiral expansion.

If we choose for the gluon condensate
the value \GG = (360 MeV)$^4$ 
(which gives the best fit of ${\cal A}(K^+\to\pi^+\pi^0)$), we obtain 
at $\mu = 0.8$ GeV
\beq
\widehat B_K \simeq 0.33\, .
\label{BKp2}
\eeq
This value includes the non-factorizable effects of gluon condensate 
corrections, which play a crucial role in the fit of the
$\Delta I = 3/2$ amplitude in $K\to \pi\pi$ decays. 

The value in \eq{BKp2} represents the starting point of our analysis,
to which we will add the effect of chiral loop contributions to 
the $\Delta S=2$ matrix element.   

\section{Computing $B_{K}$}

In this section we will extend the techniques that we have developed for
$\Delta S = 1$ weak processes in ref. \cite{HME},
by using the $\chi$QM to construct the $\Delta S =2$\ weak
chiral lagrangian.

\subsection{The leading chiral coefficient in the chiral quark model}

At the leading $O(p^2)$ order in the chiral expansion, 
the strong interaction 
between the $SU(3)$ Goldstone bosons is described by the following 
effective lagrangian~\cite{GassLeut}
\beq
{\cal L}_{\rm strong}^{(2)} =
\frac{f^2}{4} \Tr \left( D_\mu \Sigma D^\mu \Sigma^{\dag} \right)
+ \frac{f^2}{2}
B_0 \Tr \left( {\cal M} \Sigma^{\dag} +  \Sigma {\cal M}^{\dag} \right) 
\, ,
\label{strong}
\eeq
where$\cal M$ is the mass matrix of the three light quarks (u,d and s) and
$\Sigma$ is defined as
\beq
\Sigma \equiv \exp \left( \frac{2i}{f} \,\Pi (x)  \right) \, ,
\qquad \qquad
 \Pi (x) = \sum_{a=1..8} \lambda_a \pi^a (x) /2 \,.
\label{dfs2}
\eeq
To the same order, the $\Delta S=2$ weak chiral  lagrangian is given by:
\beq
{\cal L}^{(2)}_{\Delta S = 2}  = 
C(Q_{S2})
 \Tr \left( \lambda^3_2  \Sigma  D_{\mu}\Sigma^{\dag} \right)
  \Tr \left( \lambda^3_2  \Sigma  D^{\mu}\Sigma^{\dag} \right)
\, .
\label{ds2}
\eeq 
In \eq{ds2} $D_{\mu}$ indicates the covariant derivative with respect
to any external field, while $\lambda^3_2$ is a combination of the $SU(3)$ 
Gell-Mann matrices which acts in the flavor space causing a 
transition from a $d$-quark to an $s$-quark:
$(\lambda_2^3)_{lk} =\delta_{3l} \delta_{2k}$.

\begin{figure}[t]
\epsfxsize=7cm
\centerline{\epsfbox{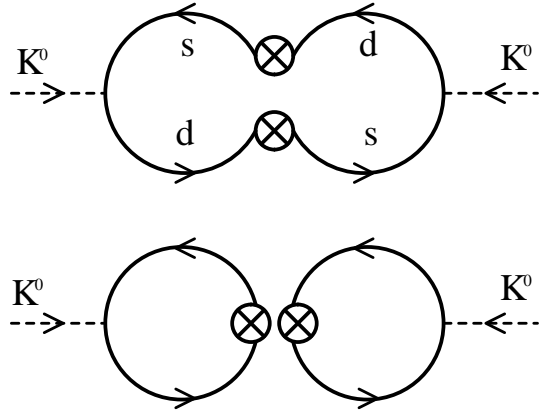}}
\caption{The two configurations relevant to the determination of the
chiral coefficient $C(Q_{S2})$. The crossed circles stand for the
insertion of the currents of the local $\Delta S = 2$
four-quark operator.}
\end{figure}

$C(Q_{S2})$ is the chiral coefficient, which we determine
by comparison with the $\chi$QM calculation. 
Two configurations contribute to the determination of 
this coefficient at $O(N_c)$,
as shown in Fig. 1.

In both HV and NDR schemes we find (for convenience we do not
write the overall $\Delta S = 2$ Wilson coefficient given in \eq{lags2}) 
\beq
C(Q_{S2}) = -\frac{f^4}{4} \left(1 + \frac{1}{N_c} \right) 
\, .
\label{cbs}
\eeq

An important correction to \eq{cbs}
arises by considering the propagation of quarks
in an external gluon field.
The effects of non-perturbative gluonic 
corrections have been first studied in~\cite{PdeR}. 

In the case of $K^0 \to \bar{K}^0$ transition the 
relevant gluonic corrections  
are given by the diagrams of Fig. 2.
\begin{figure}[t]
\epsfxsize=14cm
\centerline{\epsfbox{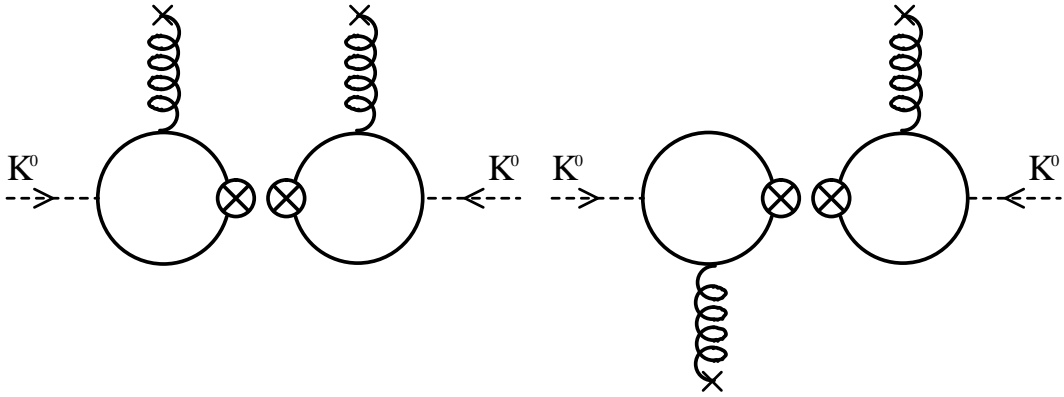}}
\caption{Same as Fig. 1 with the inclusion of the gluon corrections.}
\end{figure}
By including
gluonic condensate corrections, \eq{cbs} becomes
\beq
C(Q_{S2}) = -\frac{f^4}{4} \left[ 1 + \frac{1}{N_c} \left(1 
-\delta_{\langle GG \rangle} \right) \right] 
\, ,
\label{cbs2}
\eeq
where $\delta_{\langle GG \rangle}$ is given by \eq{gg}.

By using the definition given in \eq{bk} and computing  at 
leading order $\langle \bar{K}^0|Q_{S2}|K^0 \rangle$, we obtain the following 
expression for $B_K(\mu)$:
\beq
B_K(\mu) = \frac{3}{4} 
\left[ 1 + \frac{1}{N_c} \left(1 
-\delta_{\langle GG \rangle}\right)\right] \frac{f^2}{f_K^2}
\, .
\label{bklong}
\eeq
At this stage of the computation, $B_K (\mu)$ does not exhibit 
yet an explicit dependence on $\mu$. In our approach the scale dependence 
arises from meson-loop corrections.

If we take $f = f_ \pi$ in \eq{bklong} we recover  \eq{BKp2}, as
it should be.

Taking $f=f_K$ and 
$\delta_{\langle GG \rangle} = 0$, \eq{bklong}
reproduces the result obtained in the $1/N_c$ approach.

\subsection{One-loop renormalization of the $\Delta S = 2$ transition}

So far we have ignored chiral-loop corrections to the evaluation of $B_K$.
The introduction of these contributions gives a 
long-distance $\mu$ dependence to 
$B_K(\mu)$.  

In conventional chiral perturbation theory the scale dependence of  
meson loops renormalization is canceled by construction by the 
$O(p^4)$ counterterms in the chiral lagrangian.
In our approach, on the contrary, the tree-level 
counterterms are $\mu$ 
independent and the scale dependence introduced in the hadronic
matrix elements via the meson 
loops, evaluated in dimensional regularization
with the standard minimal subtraction, is matched with the scale dependence 
of  the Wilson coefficients. 

\begin{figure}[t]
\epsfxsize=8cm
\centerline{\epsfbox{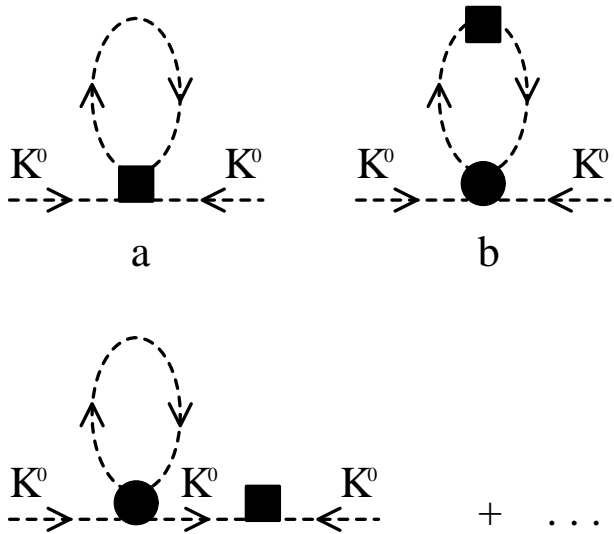}}
\caption{One-loop chiral corrections to the kaon mass matrix.
The black box and circle indicate the insertion of the weak $\Delta S = 2$
and strong chiral hamiltonians respectively. 
The octet mesons $K$, $\pi$, $\eta$ are exchanged in the loop.}
\end{figure}

The diagrams relevant to the present case
are depicted in Fig. 3.
The diagram in Fig. 3(a) contains only a four-meson weak vertex
of the $\Delta S = 2$ chiral Lagrangian, while the diagram 
in Fig. 3(b) contains two vertices, one of which is a four-meson 
strong vertex and the other is a two-meson weak vertex.
Another class of diagrams, which are induced by wave function renormalization,
is shown below 3(a,b).

A direct calculation yields:
\bea 
\langle \bar{K^0} |Q_{S2}| K^{0} \rangle |_{tree} &=&
C(Q_{S2}) \left( \frac{- 4 m_K^2}{f^2}\right)\\ 
\label{ctree}
\nnu
\langle \bar{K^0} |Q_{S2}| K^{0} \rangle |_{1a}\ \ &=&
\ C(Q_{S2})\ \frac{1}{f^4}  \left[
3 I_4 \left(m_{\eta}^2\right) + I_4 \left(m_{\pi}^2\right) 
+ 7 m_K^2 I_2  \left(m_{\eta}^2\right) \right.\\
 & & + \left. 5 m_K^2 I_2 \left(m_{\pi}^2\right)  
+\frac{8}{3} m_K^2 I_2 \left(m_K^2\right)
\right]\\
\label{ca}
\nnu
\langle \bar{K^0} |Q_{S2}| K^{0} \rangle |_{1b}\ \ & =&
\frac{4}{3} \ C(Q_{S2})\ \frac{1}{f^4} \left[ I_4 \left(m_K^2\right)
+3 m_K^2 I_2 \left(m_K^2 \right) \right.\\
 &  & + \left.  3 m_K^4  I_3 \left(m_K^2\right)\right]\\
\label{cb}
\nnu
\langle \bar{K^0} |Q_{S2}| K^{0} \rangle |_{wf} &=&
-4 \ C(Q_{S2})\ \frac{m_K^2}{f^4} 
\left[ \frac{1}{4} I_2 \left(m_{\pi}^2\right)
			 +\frac{1}{4} I_2 \left(m_{\eta}^2\right)\right.\\
 & &+ \left. \frac{1}{2} I_2 \left(m_K^2\right)\right] \, ,
\label{wf}
\eea
where $C(Q_{S2})$ is the chiral coefficient given by \eq{cbs2}, and
\beq
I_2\left(m_i^2\right) = \frac{i}{\left(2\pi\right)^4} \int \frac{1}
{\left(q^2-m_i^2\right)}d^4 q= 
\frac{1}{16\pi^2}\, m_i^2 \left(\ln\, 
\frac{m_i^2}{\mu^2} - 1\right)
\label{i2}
\eeq
\beq
I_3\left(m_i^2\right) =\frac{i}{\left(2\pi\right)^4} \int \frac{1}
{\left(q^2-m_i^2\right)^2} d^4 q= \frac{1}{16\pi^2} 
\ln\, \left(\frac{m_i^2} { \mu^2}\right) 
\label{i3}
\eeq
\beq
I_4\left(m_i^2\right) =\frac{i}{\left(2\pi\right)^4} \int \frac{q^2}
{\left(q^2-m_i^2\right)} d^4 q=
\frac{1}{16\pi^2} \, m_i^4 \left(\ln\, \frac{m_i^2}{\mu^2} - 1 
\right)
\, .
\label{i4} 
\eeq

We also have to consider
the meson decay constant renormalization, that is the one-loop determination
of $f$ in terms of $f_K$:
\beq
f = f_K \left\{ 1 + \frac{1}{2 f^2} \left[ \frac{3}{4} I_2 (m_\pi^2)
+ \frac{3}{2} I_2 (m_K^2) +  \frac{3}{4} I_2 (m_\eta^2) \right] + ... \right\}
\label{delta-f} \ ,
\eeq
where the dots represent contributions of the $O(p^4)$ chiral lagrangian.
This renormalization introduces 
chiral corrections which cancel some of the contributions coming from the 
meson loops in Fig. 3.
Notice that the divergent integrals in \eqs{i2}{i4} are minimally subtracted, 
while the usual chiral expansion prescription keeps
only the chiral logarithms. We will discuss the implications for the
$O(p^4)$ counterterms in a forthcoming publication \cite {p4}.

The expression for the $\Delta S = 2$ amplitude, comprehensive of chiral
loops, wave function and kaon decay constant 
renormalizations can be written as
\bea
\nnu
\langle \bar{K^0} |Q_{S2}| K^{0} \rangle &=&
\langle \bar{K^0} |Q_{S2}| K^{0} \rangle |_{tree} 
\left(1 + 2\frac{f-f_K}{f_K}
\right)
 \\ 
 & & +\ \langle \bar{K^0} |Q_{S2}| K^{0} \rangle |_{1a+1b+wf} 
 \, ,
\label{BKp4}
\eea
where the chiral lagrangian coefficient $f$ in the term $\langle \bar{K^0} 
|Q_{S2}| K^{0} \rangle |_{tree}\propto f^2$
is replaced by the
physical decay constant $f_K$ (all other terms are independent on $f$). 
As it was noted by Bruno in ref. \cite{Bruno},
this replacement corresponds to resumming all orders of the chiral expansion
in the factorizable component of the amplitude.

An approach similar to the one we are adopting here has been followed 
in ref. \cite{BBG} in the framework of a cut-off regularization of the
chiral loops. 
It is important to stress that we have chosen to regularize the divergent 
integrals appearing in the meson loops by using dimensional regularization (as
we have already done in~\cite{HME,II,III}).
This choice is motivated by consistency  with the 
short distance calculation of the Wilson coefficients, which is 
performed using the same regularization.

In order to show the impact of chiral loops on the
$K^0$-$\bar K^0$ amplitude, we find convenient
to factorize the tree level contribution
in terms of  the relevant parameters, while giving 
the corresponding loop renormalization as a numerical 
coefficient with an explicit $\mu$ dependence.
The values of the meson masses and other input variables are those
given in Table 2. 

We thus find:
\beq
\langle \bar{K^0} |Q_{S2}| K^{0} \rangle = 
m_K^2 f_K^2 \left[ 1 + \frac{1}{N_c} \left(1 -\delta_{\langle GG \rangle} 
\right) \right] \left(1+ 0.728 + 0.372 \  \ln \mu^2 \right) \ , 
\label{ren}
\eeq
where $\mu$ is taken in units of GeV.
From \eq{ren} we obtain the final result for $B_K(\mu)$, 
inclusive of the effects of meson loops, wave function and kaon decay
constant renormalization:
\beq
B_K(\mu) =\frac{3}{4} \left[ 1 + \frac{1}{N_c} \left(1 -\delta_{\langle GG 
\rangle} \right) \right] \left(1+0.728+0.372 \, \ln \mu^2 \right) \ .
\label{bkn}
\eeq

The scale dependence of the hadronic matrix element interferes 
constructively with that of $b(\mu)$ in \eq{wc}.
Nevertheless, the overall scale dependence of $\widehat B_K$
remains below 20\% in the range between 0.8 and 1 GeV.

\section{Numerical Analysis}

We now have all the ingredients necessary to make a detailed analysis of the 
values of the parameter $\widehat B_K$, where $B_K(\mu)$ and $b(\mu)$ are 
given by \eq{bkn} and \eq{wc}, respectively.
The final result depends on the values of 
the gluon condensate \GG 
entering in the determination of the gluon corrections to $B_K(\mu)$ 
and of $\Lambda_{\rm QCD}^{(4)}$ 
which determines the value of the QCD coupling constant 
$\alpha_S$, and consequently of $b(\mu)$ .

\subsection{Input Parameters}

A relevant input parameter in our present analysis is
the gluon condensate.
We choose for this quantity the value that gives within a $30$\% error
a fit of the $\Delta I=3/2$ $K^+\to\pi^+\pi^0$ amplitude:
\beq
\left\langle \frac{\alpha_S}{\pi} GG \right\rangle = 
\left(360 \pm 15 \: \mbox{MeV}\right)^4 \, . 
\label{Gscale}
\eeq
Although this value fits leading order QCD sum rule determinations, 
the relation between the chiral quark model gluon condensate 
and other estimates (e.g.: QCD sum rules) is rather unclear (only low frequency
modes of the gluon fields are included in \eq{Gscale}).
We stress that our approach is in this respect
purely phenomenological: we
consider the quark and gluon condensates as parameters of the model to
be consistently determined by comparison with known observables.
A redundant set of determinations provide the basis for a predictive 
framework.  
\begin{table}[t]
\begin{center}
\begin{tabular}{|c|c|}
\hline
{\rm parameter} & {\rm value} \\
\hline
$f_\pi = f_{\pi^+}$  &  92.4  MeV \\
$f_K = f_{K^+}$ & 113 MeV \\
$m_\pi = (m_{\pi^+} + m_{\pi^0})/2 $ & 138 MeV \\
$m_K = m_{K^0}$ &  498 MeV \\
$m_\eta$ & 548 MeV \\
$\Lambda_{\rm QCD}^{(4)}$ & $350 \pm 100$ MeV \\
\hline
\end{tabular}
\end{center}
\caption{Table of the numerical values used for the input parameters.}
\end{table}
A word of caution concerning the renormalization
prescription of the chiral lagrangian
parameter $f$ in the amplitudes:
in refs. \cite{HME,II,III}
we have included the one-loop renormalization of $1/f^3$ 
in the $K\to\pi\pi$ tree level chiral amplitudes.  
From now on we include in the counting of powers of $f$
also the $f$ dependence of the chiral coefficient computed in the 
$\chi$QM.
The numerical consequences of this change in prescription are that the best
fit of the $\Delta I = 1/2$ rule leads to a central value of the gluon 
condensate $\langle\alpha_s G G/\pi \rangle =$ (360 MeV)$^4$, 
slightly smaller than that obtained in \cite{II}, namely (372 MeV)$^4$,
and to a central value of the quark condensate of $(- 280 MeV)^3$,
slightly larger than $(- 271 MeV)^3$,
quoted as the best fit in ref. \cite{II}. 
Since our present results 
depend very little on the quark condensate we keep it
fixed at the value $\vev{\bar q q} = -$ (280 MeV)$^3$, while varying
the gluon condensate in the range of \eq{Gscale}.

The present analysis, as our previous ones, includes $O(p^2)$ chiral
coefficients and the effects of chiral loops. A complete $O(p^4)$
calculation is under way \cite{p4}. Preliminary results show that
the best fit of the $\Delta I = 1/2$ rule is obtained for lower values
of the quark and gluon condensates. 

Another input parameter which is important for the determination 
of $\widehat B_K$, 
is the QCD running coupling constant $\alpha_s$ entering in 
the computation of the short distance factor $b(\mu)$.
In our numerical estimates we use for
$\alpha_s$ the range of eq.(\ref{alfar}), corresponding to the values of 
$\Lambda_{\rm QCD}^{(4)}$ given by (\ref{lambdone}).
The values of this and other input parameters are listed in Table 2.

\subsection {Numerical results for $\widehat B_K$}

Our numerical estimate of the parameter $\widehat 
B_K$ is summarized in Table 3, in which we have
fixed \GG
to the central value of  \eq{Gscale}
and we have examined two extreme values of the matching scale $\mu$ in 
both schemes HV and NDR. 
The three parts of the table show the dependence on the QCD scale 
parameter $\Lambda_{\rm QCD}^{(4)}$. 

From Table 3 we obtain the ranges 
$0.79 \leq \widehat B_K \leq 1.0$ in NDR and $0.69 \leq \widehat B_K 
\leq 0.97$ in HV scheme.
\begin{table}
\begin{footnotesize}
\begin{center}
\begin{tabular}{|c||c|c||c|c|}
\hline
\multicolumn{5}{|c|}{$\Lambda^{(4)}_{\rm QCD}$ = 250 MeV}\\
\hline 
 & \multicolumn{2}{c||}{$\mu = 0.8$ \mbox{GeV}} 
 & \multicolumn{2}{c|}{$\mu = 1$ \mbox{GeV}} \\
\hline
       & NDR & HV &  NDR & HV \\
\hline
$b\left(\mu\right)$  & 1.25 & 1.19 &  1.30 & 1.24  \\
\hline

$\widehat B_K$  & 0.88 & 0.84 &  1.01 & 0.97 \\
\hline
$\Delta_{\gamma_5} \widehat B_K  $ & \multicolumn{2}{c||}{$ 5.2 \%$ } &
\multicolumn{2}{c|}{$ 4.2 \%$ }\\
\hline
$\Delta_\mu \widehat B_K $ & \multicolumn{4}{c|}{$14\% - 15\%$ } \\
\hline
$\Delta_\mu b\left(\mu\right) $ & \multicolumn{4}{c|}{$4\% -5\%$ } \\
\hline
\hline
\multicolumn{5}{|c|}{$\Lambda^{(4)}_{\rm QCD}$ = 350 MeV}\\
\hline
 & \multicolumn{2}{c||}{$\mu = 0.8$ GeV} 
 & \multicolumn{2}{c|}{$\mu = 1$ GeV} \\
\hline
       & NDR & HV & NDR & HV \\
\hline
$b\left(\mu\right)$  & 1.17 & 1.08 &  1.23 & 1.17  \\
\hline
$\widehat B_K$  & 0.83 & 0.77  & 0.96 & 0.91 \\
\hline
$\Delta_{\gamma_5} \widehat B_K  $ & 
\multicolumn{2}{c||}{$ 8 \%$ } &
\multicolumn{2}{c|}{$ 5.7 \%$ }\\
\hline
$\Delta_\mu \widehat B_K $ & \multicolumn{4}{c|}{$15\% - 17\%$ } \\
\hline
$\Delta_\mu b\left(\mu\right) $ & \multicolumn{4}{c|}{$5\% -7\%$ } \\
\hline
\hline
\multicolumn{5}{|c|}{$\Lambda^{(4)}_{\rm QCD}$ = 450 MeV}\\
\hline
 & \multicolumn{2}{c||}{$\mu = 0.8$ GeV} 
 & \multicolumn{2}{c|}{$\mu = 1$ GeV} \\
\hline
       & NDR & HV &  NDR & HV \\
\hline
$b\left(\mu\right)$  & 1.12 & 0.97 &  1.18 & 1.09  \\
\hline
$\widehat B_K$  & 0.79 & 0.69 &  0.92 & 0.85 \\
\hline
$\Delta_{\gamma_5} \widehat B_K  $ & \multicolumn{2}{c||}{$ 14.1 \%$ } &
\multicolumn{2}{c|}{$ 7.9 \%$ }\\
\hline
$\Delta_\mu \widehat B_K $ & \multicolumn{4}{c|}{$15\% - 21\%$ } \\
\hline
$\Delta_\mu b\left(\mu\right) $ & \multicolumn{4}{c|}{$5\% -11\%$ } \\
\hline
\end{tabular}
\end{center}
\end{footnotesize}
\caption{Matching scale and $\gamma_5$ scheme dependence of $\widehat B_K$
in the $\chi$QM with NLO Wilson coefficients, for various values
of $\Lambda^{(4)}_{\rm QCD}$.
We take for the gluon condensate the
value $\vev{\alpha_s GG / \pi}$ = (360 MeV)$^4$, preferred by the fit of
$\Gamma(K^+\to\pi^+\pi^0)$.}
\end{table}

The quantities 
$\Delta_{\gamma_5} \widehat B_K$ and $\Delta_{\mu} \widehat B_K$
measure the size of the $\gamma_5-$scheme and $\mu-$ dependences respectively, 
\bea
\Delta_{\gamma_5} \widehat B_K &=& 2 \left|\, \frac{\widehat B_K|_{\rm HV}-
\widehat B_K|_{\rm NDR}}{\widehat{B}_K|_{\rm HV}
+\widehat B_K|_{\rm NDR}}\, \right|\\
\Delta_{\mu} \widehat B_K &=& 2 \left|\, \frac{\widehat B_K(1\: \mbox{GeV})-
\widehat B_K(0.8\: \mbox{GeV})}{\widehat B_K(1\: \mbox{GeV})+\widehat B_K
(0.8\:\mbox{GeV})}
\, \right|
\, .
\eea 

The scale dependence $\Delta_{\mu} \widehat B_K$ is near to $20\%$ in both 
schemes and it is mainly due to the effect of meson loops renormalization.
As a matter of fact the final $\mu$ dependence is larger
 than the one originally 
present in the coefficient $b(\mu)$, which is less than $10\%$. 
Nevertheless the fact that the scale dependence is at most $20\%$ 
makes us confident on the stability of our results.
and allows us to choose 
$\mu=0.8$ GeV as the best compromise 
between the upper limit of validity of chiral perturbation theory, used to 
compute $B_K(\mu)$, and the lowest scale for
perturbative calculations, needed to 
obtain the short distance coefficient $b(\mu)$.

The scheme dependence of our result
is entirely due to $b(\mu)$, since the hadronic matrix element 
does not exhibit any scheme dependence.
At any rate $\Delta_{\gamma_5} \widehat B_K$ is below
$10\%$ for all values of $\mu$ in the given range. 

Finally, a few words on the dependence of our results on the 
value chosen for the gluon condensate. In Fig. 4 
we show $\widehat B_K$ 
as a function of the gluon condensate, for our preferred matching scale 
$\mu=0.8$ GeV, and $\Lambda^{(4)}_{\rm QCD}= 350$ MeV.

\begin{figure}[t]
\epsfxsize=12cm
\centerline{\epsfbox{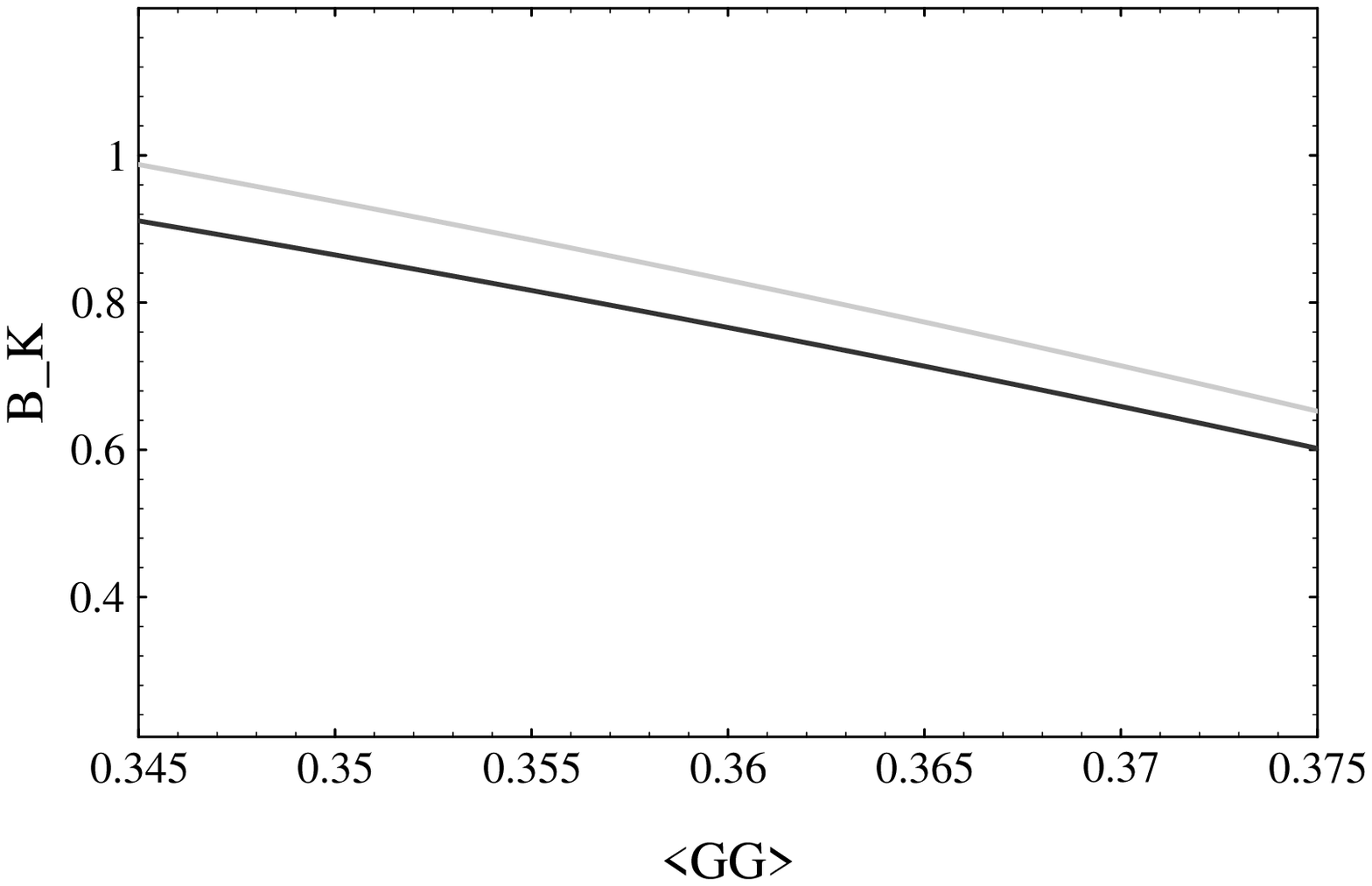}}
\caption{The $\widehat B_K$ parameter is shown as a function of the gluon 
condensate for $\Lambda^{(4)}_{\rm QCD} = 350$ MeV and $\mu=0.8$ GeV.
We denote by $\vev{GG}$ 
the quantity $\vev{\alpha_s GG/\pi}^{1/4}$ in units of GeV.
The dark and grey lines represent the HV and NDR results respectively.}
\end{figure}

It appears that $\widehat B_K$ is a sensitive decreasing
function of the gluon condensate. 
For our values of \GG  the term $\delta_{\langle GG \rangle}$ 
in \eq{bkn} is greater than
1, thus changing the sign of the $1/N_c$ contribution and determining a 
reduction of the final result for $\widehat B_K$. 
By varying also
the value of the gluon condensate in the range of \eq{Gscale}
we obtain the overall range 
\beq
0.54 \leq \widehat B_K \leq 1.2 \ .
\label{unbasiedbk}
\eeq
which represents our conservative prediction for $B_K$.
The central value $\widehat B_K = 0.87$, quoted in the abstact, is obtained
by taking all input parameters at their central values. It represents a large
renormalization with respect to the initial value given by \eq{BKp2}.
In this respect, improving to $O(p^4)$ the chiral expansion, albeit
challenging, might be needed in order 
to assess the degree of stability of the result.

\section{$K_L$-$K_S$ mass difference}

We apply the results of the previous section to the study
of the $K^0_L$-$K^0_S$ mass difference $\Delta M_{LS}$.
The full $\Delta M_{LS}$ can be split into short-  and long-distance
components as
\beq
\Delta M_{LS} = \Delta M_{SD} + \Delta M_{LD} \, .
\eeq

Notice that the ``short-distance'' component $\Delta M_{SD}$, 
generated by the lagrangian in \eq{lags2}, 
contains the hadronic parameter $\widehat B_K$.
The value of $\widehat{B}_K$ estimated in the 
previous part of the paper completes the determination of
the box (or short-distance) component of $\Delta M_{LS}$. 

We address now the issue of the evaluation of the genuine long-distance
contribution $\Delta M_{LD}$. We will do it consistently with
the evaluation of $B_K$.
The interesting question is whether
the interplay between $\Delta M_{SD}$ and 
$\Delta M_{LD}$ reproduces the
observed value $\Delta M^{exp}_{LS}$.

\subsection{Long distance $\Delta S = 1$ induced contributions}

Many attempts have been made to estimate $\Delta M_{LD}$
\cite{DGH1}--\cite{BGK} by means of various
 techniques like chiral symmetry, dispersion relation
with experimental data of $s-$wave $\pi\pi$ scattering, 
leading to a variety of numerical results. 

Our aim is to give a consistent
estimate of $\Delta M_{LD}$ based on the 
$\chi$QM approach. 

As already mentioned, the $K^0$-$\bar K^0$ mass difference
receives contributions from the exchange of the $SU(3)$ meson field octet
(we leave aside in this analysis the contribution of $\eta'$)
via the double insertion of the $\Delta S=1$ chiral vertices.

The complete 
bosonization of the $\Delta S=1$ lagrangian of \eq{ham}
can be found in ref.~\cite{ML}. 
Here we just quote the result for 
the operators $Q_{1-6}$, which turn out to be relevant 
for the calculation of 
$\Delta M_{LD}$. 
The electroweak penguins $Q_{7-10}$ 
give a negligible 
contribution (of the order of $1\%$)
due the smallness of their Wilson coefficients.

The bosonization of the relevant operators 
leads to
\bea
{\cal L}^{(2)}_{\Delta S = 1}  & = &G_{\underline{8}} (Q_{3-6}) \Tr \left( 
\lambda^3_2 D_\mu \Sigma^{\dag}
D^\mu \Sigma
\right) +  \nnu \\
 & & \: G_{LL}^a (Q_{1,2}) \, 
\Tr \left(  \lambda^3_1 \Sigma^{\dag} D_\mu \Sigma \right)
\Tr \left( \lambda^1_2 \Sigma^{\dag} D^\mu  \Sigma \right) +  \nnu \\
 & & \: G_{LL}^b (Q_{1,2})\, \Tr \left( \lambda^3_2 \Sigma^{\dag} D_\mu
\Sigma \right)
\Tr \left(  \lambda^1_1 \Sigma^{\dag} D^\mu \Sigma \right)  
\label{chi-lag} \, ,
\eea
where, as before, $\lambda^i_j$ are combinations of Gell-Mann
$SU(3)$ matrices defined by $(\lambda^i_j)_{lk} = \delta_{il}\delta_{jk}$
and $\Sigma$ is defined in \eq{dfs2}.
The covariant
derivatives in \eq{chi-lag} are taken with respect to the external
gauge fields whenever they are present.

\begin{table}
\begin{center}
\begin{tabular}{|l|l|}
\hline 
\hspace*{2.5cm} HV & \hspace{2.5cm} NDR 
\\
\hline
$ G_{LL}^a(Q_1)  =  - \frac{1}{N_c} f_\pi^4  \left( 1 - \delta_{\vev{GG}}
\right) \nnu $  & $ G_{LL}^a(Q_1) =  - \frac{1}{N_c} f_\pi^4  \left( 1 - \delta_{\vev{GG}}
\right) \nnu $
\\ 
$ G_{LL}^a(Q_2) =  - f_\pi^4  \nnu $  & $ G_{LL}^a(Q_2)  =  - f_\pi^4 $ \nnu
\\
$ G_{LL}^b(Q_1)  =  - f_\pi^4  \nnu $  & $ G_{LL}^b(Q_1) =  - f_\pi^4  \nnu $
\\
$ G_{LL}^b(Q_2)  = - \frac{1}{N_c} f_\pi^4 \left( 1 - \delta_{\vev{GG}} \right)
  \nnu $ &  
$ G_{LL}^b(Q_2)  = - \frac{1}{N_c} f_\pi^4 \left( 1 - \delta_{\vev{GG}} \right)
  \nnu $
\\ 
 $ G_{\underline{8}} (Q_3)  =    f_\pi^4 \frac{1}{N_c} \left( 1 -
\delta_{\vev{GG}}
\right) \nnu $ &  $ G_{\underline{8}} (Q_3)  =   f_\pi^4 \frac{1}{N_c} \left( 1 -
\delta_{\vev{GG}}
- 6 \frac{M^2}{\Lambda_\chi^2}
\right) \nnu $ 
\\
$ G_{\underline{8}} (Q_4)  =    f_\pi^4 \nnu $ & $G_{\underline{8}} (Q_4)  =   f_\pi^4 \left( 1 - 6 \frac{M^2}{\Lambda_\chi^2}
\right) \nnu $ 
\\ 
$ G_{\underline{8}} (Q_5)  =   \frac{2}{N_c} \,
\frac{\vev{\bar{q}q}}{M} f_\pi^2 \, \left( 1 - 6\,
\frac{M^2}{\Lambda_{\chi}^2} \right) \nnu $  & $ G_{\underline{8}} (Q_5) =  \frac{2}{N_c} \,
\frac{\vev{\bar{q}q}}{M} f_\pi^2 \, \left( 1 - 9\,
\frac{M^2}{\Lambda_{\chi}^2} \right)\nnu $
\\
 $ G_{\underline{8}} (Q_6)  =  2 \,
\frac{\vev{\bar{q}q}}{M} f_\pi^2 \, \left( 1 - 6\,
\frac{M^2}{\Lambda_{\chi}^2} \right)  \nnu $ & $ G_{\underline{8}} (Q_6) =  2 \,
\frac{\vev{\bar{q}q}}{M} f_\pi^2 \, \left( 1 - 9\,
\frac{M^2}{\Lambda_{\chi}^2} \right)  \nnu $
\\
\hline
\end{tabular}
\end{center}
\caption{Values of the relevant $\Delta S = 1$ 
weak chiral coefficients for two different 
regularization schemes: HV and NDR. The inclusion of the Wilson 
coefficients of the effective quark operators $Q_i$ is understood.}
\label{Gchi}
\end{table}

The notation for the chiral coefficients $G_{\underline{8}} (Q_{3-6})$, 
$G_{LL}^a(Q_{1,2})$ and $G_{LL}^b (Q_{1,2})$
reminds us their chiral properties: 
$G_{\underline{8}}$ represents the $(\underline{8}_L \times 
\underline{1}_R)$ part of the interaction induced in QCD by the
gluonic penguins, while the two terms proportional to $G_{LL}^a$ 
and $G_{LL}^b$ 
are admixtures of the $(\underline{27}_L \times \underline{1}_R)$ and the
$(\underline{8}_L \times \underline{1}_R)$ part of the interaction,
induced by left-handed current-current operators. 
These coefficient have been evaluated in two different 
schemes of regularization HV and NDR, and the results are given in Table 4. 
In this table   
$M$ is the constituent quark
mass, that, consistently with previous analyses, we take at 220 MeV and
$\Lambda_\chi$ is the chiral symmetry breaking scale ($\simeq$ 1 GeV).

The diagrams relevant to 
the evaluation of the long-distance contribution $\Delta M_{LD}$ 
arise via one-particle and two-particle 
intermediate states 
(three-particle intermediate states have been shown not to
give significant contributions~\cite{CN}).
They contain two weak vertices, among those proportional to 
$G_{\underline{8}}$, $G_{LL}^a$ and $G_{LL}^b$.
Therefore we have to consider all the possible combinations:
$ G_{LL}^a  G_{LL}^a$, $G_{LL}^b G_{LL}^b$, 
$G_{\underline{8}} G_{\underline{8}}
$, $G_{LL}^a G_{LL}^b$, $G_{\underline{8}} G_{LL}^a$, 
$G_{\underline{8}} G_{LL}^b$. 

Using the Feynman rules reported in appendix A, 
it is found that the single particle intermediate state 
contribution give a result proportional to
$ (4 m_K^2 - m_{\pi}^2 - 3 m_{\eta}^2$) 
which vanishes ~\cite{DGH1} by the Gell-Mann-Okubo relation.

\begin{figure}[t]
\epsfxsize=11cm
\centerline{\epsfbox{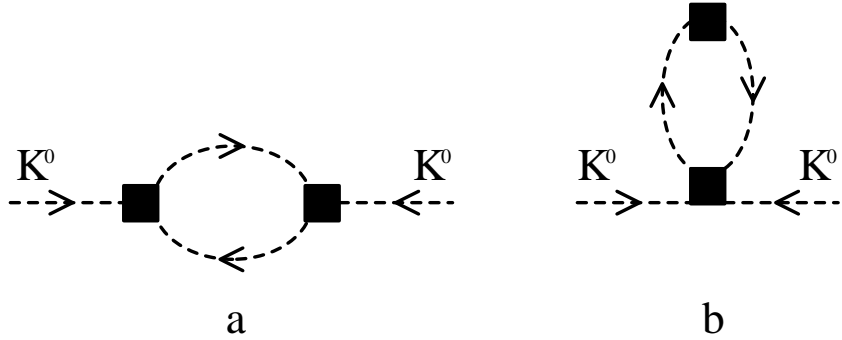}}
\caption{One-loop long-distance contributions to the $K^0-\bar K^0$ mixing
induced by the $\Delta S =1$ weak hamiltonian.
The black box represents the insertion of the $\Delta S = 1$ chiral
interactions.
The octet mesons $K$, $\pi$, $\eta$ are exchanged in the loop.}
\end{figure}

A non-vanishing contribution is obtained from the two particle intermediate 
states, which corresponds to the double insertion of the $\Delta S=1$ chiral 
lagrangian as depicted in Fig. 5(a) and 5(b). 
To our knowledge, 
the relevance of the diagrams of the type (b) (tadpole diagrams)
was first pointed out in ref.~\cite{KUR}.

The calculation is lengthy and the details can be found 
in refs.~\cite{PhD}.   
In evaluating the loop integrals, 
we use dimensional regularization and modified minimal subtraction.

\subsection{$\Delta S = 1$ Wilson coefficients}

In Table 5 we report the Wilson coefficients of the first six
operators at the scale $\mu = 0.8$ GeV in the NDR and HV 
$\gamma_5$-schemes, respectively.
Since $\Re \tau$  in \eq{ham} is of $O(10^{-3})$, 
the $K^0\leftrightarrow \bar{K}^0$
transition is  controlled by the  coefficients $z_i$, which do not depend
on $m_t$.

\begin{table}
\begin{center}
\begin{tabular}{|c|r r||r r||r r|}
\hline
$\Lambda_{QCD}^{(4)}$ & \multicolumn{2}{c||}{ 250 MeV }
                      & \multicolumn{2}{c||}{ 350 MeV } 
                      & \multicolumn{2}{c| }{ 450 MeV } \\
\hline
$\alpha_s(m_Z)_{\overline{MS}}$ 
                      & \multicolumn{2}{c||}{ 0.113 }
                      & \multicolumn{2}{c||}{ 0.119 } 
                      & \multicolumn{2}{c| }{ 0.125 } \\
\hline \hline
\multicolumn{7}{|c|}{NDR}\\
\hline
$z_1$&$(0.0503)$&$-0.524$&$(0.0533)$&$-0.663$&$(0.0557)$&$-0.781$ \\
\hline
$z_2$&$(0.982)$&$1.29$&$(0.981)$&$1.39$&$(0.980)$&$1.48$ \\
\hline
$z_3$&$$&$0.0180$&$  $&$0.0360$&$ $&$0.0870$ \\
\hline
$z_4$&$$&$-0.0471$&$  $&$-0.0852$&$ $&$-0.182$ \\
\hline
$z_5$&$$&$0.0085$&$  $&$0.0077$&$ $&$-0.0129$ \\
\hline
$z_6$&$$&$-0.0495$&$  $&$-0.0947$&$ $&$-0.226$ \\
\hline
\hline
\multicolumn{7}{|c|}{HV}\\
\hline
$z_1$&$(0.0320)$&$-0.657$&$(0.0339)$&$-0.910$&$(0.0355)$&$-1.36$ \\
\hline
$z_2$&$(0.988)$&$1.38$&$(0.987)$&$1.58$&$(0.987)$&$1.96$ \\
\hline
$z_3$&$$&$0.0137$&$  $&$0.0301$&$ $&$0.0798$ \\
\hline
$z_4$&$$&$-0.0292$&$  $&$-0.0540$&$ $&$-0.115$ \\
\hline
$z_5$&$$&$0.0070$&$  $&$0.0100$&$ $&$0.0123$ \\
\hline
$z_6$&$$&$-0.0275$&$  $&$-0.0515$&$ $&$-0.112$ \\
\hline
\end{tabular}
\end{center}
\caption{NLO Wilson coefficients at $\mu=0.8$ GeV in the NDR and in the HV
scheme ($\alpha=1/128$).
The corresponding values at $\mu=m_W$ are given in parenthesis.
In addition one has $z_{3-6}(m_c)=0$. The coefficients
$z_i(\mu)$ do not depend on $m_t$.}
\end{table}

\subsection{$\Delta S = 2$ Wilson coefficients}

The Wilson coefficients of the $\Delta S=2$ effective quark operator
are denoted by $\eta_1$, $\eta_2$ and $\eta_3$ (see \eq{lags2}). 

The NLO calculations of $\eta_1$ and $\eta_2$ can be found 
in refs. \cite{HH1} and \cite{BJW} respectively, 
while the analogous calculation 
for $\eta_3$, which is particularly challenging,
has been performed only recently by the authors of ref. \cite{HH2}.
We have taken their results and evaluated the QCD factors
for our choice of parameters.

As an example, for 
$\Lambda^{(4)}_{\rm QCD} = 350$ MeV,
$m_t^{(pole)} = 180$ GeV, and $\mu=m_c=1.4$ GeV we find
\beq
\eta_1 = 1.36\quad \eta_2 = 0.51\quad \eta_3= 0.45
\eeq

\subsection{Numerical Analysis}
\begin{table}
\begin{footnotesize}
\begin{center}
\begin{tabular}{|c||c|c||c|c|}
\hline
\multicolumn{5}{|c|}{$\Lambda^{(4)}_{\rm QCD}$ = 250 MeV}\\
\hline 
 & \multicolumn{2}{c||}{$\mu = 0.8$ GeV} 
 & \multicolumn{2}{c|}{$\mu = 1$ GeV} \\
\hline
       & NDR & HV &  NDR & HV \\
\hline
$\Delta M_{SD}$ & 2.55 & 2.42 &  2.92 & 2.80   \\
\hline
$\Delta M_{LD}$  & -0.34  & -0.38 &  -0.35 & -0.38  \\
\hline
$D$              & -0.10  & -0.11 & -0.10  & -0.11 \\
\hline
$\Delta M^{th}/\Delta M^{exp}$ & 0.63 & 0.58 &  0.73 & 0.69 \\ 
\hline
\hline
\multicolumn{5}{|c|}{$\Lambda^{(4)}_{\rm QCD}$ = 350 MeV}\\
\hline 
 & \multicolumn{2}{c||}{$\mu = 0.8$ GeV} 
 & \multicolumn{2}{c|}{$\mu = 1$ GeV} \\
\hline
       & NDR & HV & NDR & HV \\
\hline
$\Delta M_{SD}$   & 3.07 & 2.83 &  3.56 &  3.37 \\
\hline
$\Delta M_{LD}$  & -0.54 & -0.65 &  -0.46 & -0.51 \\
\hline
$D$              & -0.15  & -0.18 & -0.13  & -0.14 \\
\hline
$\Delta M^{th}/\Delta M^{exp}$ & 0.72 & 0.62 &  0.88 &  0.81\\ 
\hline
\hline
\multicolumn{5}{|c|}{$\Lambda^{(4)}_{\rm QCD}$ = 450 MeV}\\
\hline 
 & \multicolumn{2}{c||}{$\mu = 0.8$ GeV} 
 & \multicolumn{2}{c|}{$\mu = 1$ GeV} \\
\hline
       & NDR & HV &  NDR & HV \\
\hline
$\Delta M_{SD}$   & 3.91 & 3.40 &  4.55 & 4.20  \\
\hline
$\Delta M_{LD}$  & -1.18 & -1.50 &  -0.65 & -0.75  \\
\hline
$D$              & -0.34  & -0.43 & -0.18  & -0.21 \\
\hline
$\Delta M^{th}/\Delta M^{exp}$ & 0.77 & 0.54 &  1.11 & 0.98 \\ 
\hline
\end{tabular}
\end{center}
\end{footnotesize}
\caption{Long-distance and short-distance box contributions to $\Delta M_{LS}$,
in units of $10^{-15}$ GeV, for different values
of the matching scale $\mu$ and $\Lambda^{(4)}_{\rm QCD}$ in the $\chi$QM.
We take for the gluon condensate the value $\vev{\alpha_s GG / \pi}$ = 
(360 MeV)$^4$ and for the quark condensate $\vev{\bar{q} q} = -(280$ MeV)$^3$,
which are the values preferred by the fit of the 
$\Delta I = 1/2$ selection rule at the same perturbative order. 
The ``short-distance''
component $\Delta M_{SD}$ is evaluated for a top quark pole mass of 180 GeV
and for the values of $\widehat B_K$ given in Table 3.}
\end{table}

According to Wolfenstein's 
notation \cite{wolf}, we define the parameter  $D$ which
characterizes the long distance contribution
\beq
D = \frac{\Delta M_{LD}}{\Delta M_{LS}^{exp}}\ .
\eeq

Numerical estimates of D, $\Delta M_{LD}$ and $\Delta M_{SD}$  for 
different values of $\Lambda^{(4)}_{\rm QCD}$ and of the matching scale $\mu$
are given in Table 6. In this table we have fixed the gluon condensate to our 
central value \GG = (360 MeV)$^4$ and we
have chosen for the quark condensate the value $\langle \bar{q} q \rangle =$
$-$(280 MeV)$^3$ which gives us the best fit of the $\Delta I =1/2$ 
selection rule. The ranges thus obtained are 
$0.63 \leq \Delta M_{LS}^{th}/\Delta M_{LS}^{exp} \leq 1.11$ in the NDR and
$ 0.54 \leq \Delta M_{LS}^{th}/\Delta M_{LS}^{exp} \leq  0.98$ 
in the HV. 
 
A few comments are in order.
Among the diagrams of Fig. 5(a), 
those containing two intermediate pion states dominate over kaon and 
eta exchange by about a factor of two.
The diagrams 
of Fig. 5(b) (tadpole diagrams) give a contribution
comparable in size with those of Fig. 5(a) but of the opposite sign, leading
to a small and negative $\Delta M_{LD}$ in  most  of the parameter
space.

\begin{figure}[t]
\epsfxsize=12cm
\centerline{\epsfbox{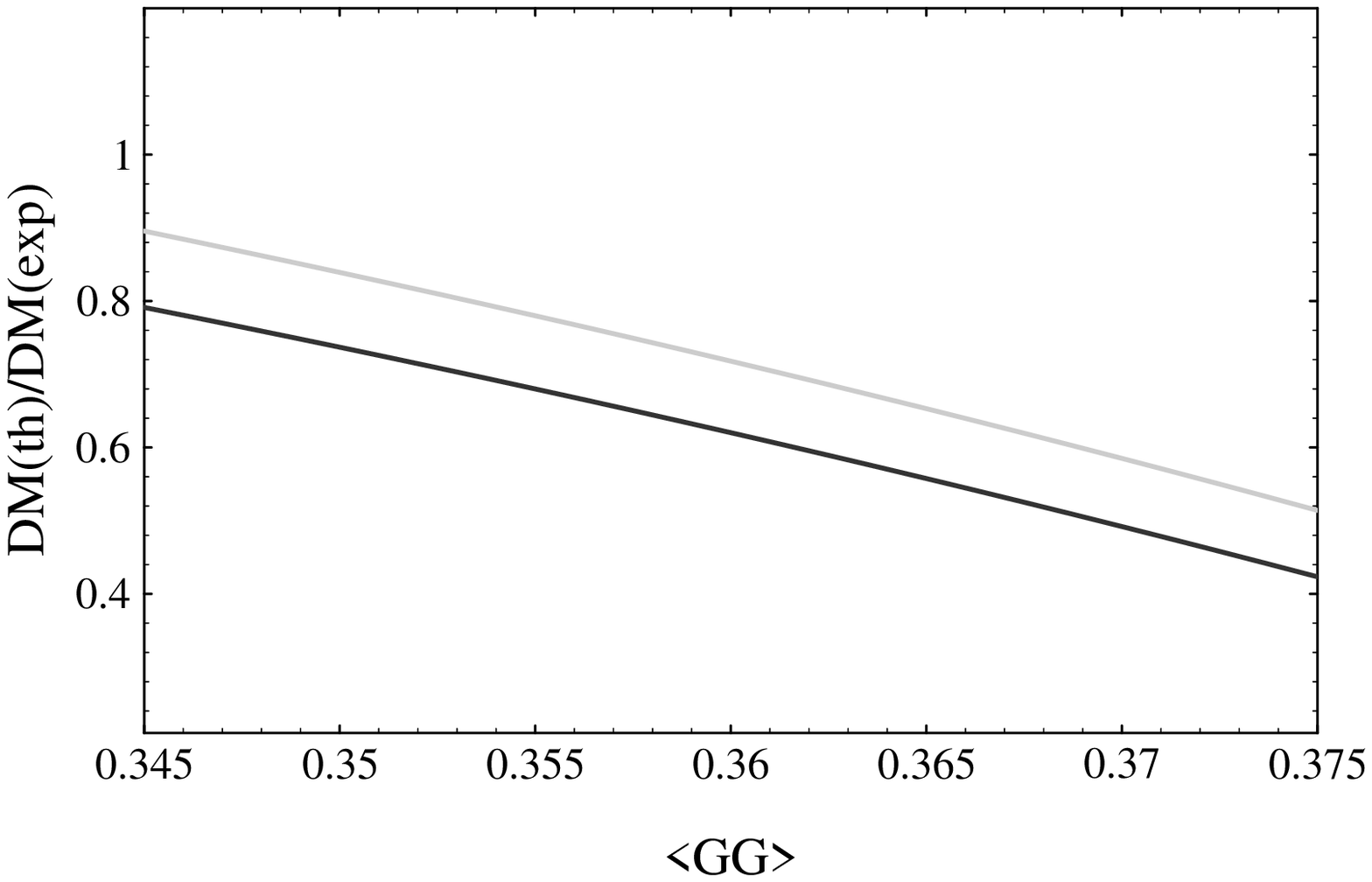}}
\caption{The ratio $(\Delta M_{SD}+\Delta M_{LD})/\Delta M_{LS}^{exp}$
is shown as a function of the gluon condensate for 
$\Lambda^{(4)}_{\rm QCD} = 350$ MeV and $\mu=0.8$ GeV.
We denote by $\vev{GG}$ 
the quantity $\vev{\alpha_s GG/\pi}^{1/4}$ in units of GeV.
The dark and grey lines represent the HV and NDR results respectively.}
\end{figure}

We disagree with
ref.~\cite{KUR} in the details of the calculation and on the
the relevant interactions.
In particular, the author of
 ref.~\cite{KUR} seems to neglect some
of the leading insertions of the operator $Q_2$.

Our result depends sensitively on the value of the gluon condensate
(the uncertainties in the short-distance coefficients
related to varying the quark thresholds---in particular
$m_c$ in the range $1.3\div 1.5$ GeV---affect $\Delta M_{LS}$
by less than 15\% for our central value of $\Lambda_{\rm QCD}$).
Fig. 6 shows the typical behavior for a choice of input parameters.
The total theoretical mass difference $\Delta M_{LS}^{th}$ is a decreasing
function of the value of the gluon condensate, analogously to the case
of $\widehat B_K$. 
If we let the value of
the gluon condensate vary in the range of \eq{Gscale}, as we did in determining
$\hat B_K$, we obtain the overall range
\beq
0.42 \leq \Delta M_{LS}^{th}/\Delta M_{LS}^{exp} \leq  1.40
\eeq
which represents our most conservative result.

The scheme dependence of the result is satisfactory ($< 20\%$)
for most of the range of $\Lambda_{\rm QCD}^{(4)}$.
Less satisfactory is the renormalization scale dependence.
$\Delta M_{SD}$ is an increasing function of $\mu$ and this behavior 
is not compensated by a corresponding 
decrease of the D parameter. This feature 
leads to a scale dependence that is about 30\%
for $\Lambda_{\rm QCD}^{(4)} \leq
350$ MeV. 
These results may indicate the need to extend the analysis to
$O(p^4)$ in the chiral lagrangian expansion. An improved
$\chi$QM calculation of the $\Delta S = 1$ and $\Delta S = 2$
chiral coefficients at $O(p^4)$ is under way.   
 
A final comment about the width difference 
$\Delta\Gamma_{LS}$ is necessary. 
A direct calculation of the absorptive component of Fig. 5(a)
gives about 1/6 of the 
experimental result. The reason is that the 
tree-level $K\to\pi\pi$ decay amplitudes do not reproduce
the measured ones. Only by replacing in the vertices of Fig. 5(a) the 
one-loop results obtained in~\cite{II}, we obtain the agreement with 
the experimental $\Delta\Gamma_{LS}$. This is equivalent
to computing directly the absorptive part of 
of Fig. 5(a) up to three loops.

\section{The Mixing Parameter $\mbox{Im}\, \lambda_t$}

A range for the KM parameter Im $\lambda_t$, which is relevant
for CP violating observables, 
can be determined from the 
experimental value of $\varepsilon$ as a function of $\widehat B_K$, $m_t$ 
and the other relevant parameters involved in the theoretical estimate.

Given $m_t$, $m_c$ and the KM parameters~\cite{PDG}  
\bea
 |V_{us}| & = & 0.2205 \pm 0.0018  \\
 |V_{cb}| & = & 0.041 \pm 0.003  \\
 |V_{ub}/V_{cb}| & = & 0.08 \pm 0.02 \, , 
\eea
we can solve the two equations
\beq
\varepsilon^{th} 
 ( \widehat B_K, |V_{cb}|, |V_{us}|, 
   \Lambda_{\rm QCD}^{(4)}, m_t, m_c, \eta,\rho ) =  
\varepsilon^{exp}  
\label{bound1}
\eeq
\beq
\eta^2 + \rho^2 =  \frac{1}{| V_{us}|^2} \left|\frac{V_{ub}}{V_{cb}}\right|^2
 \label{bound2}
\eeq
to find the allowed values of the two parameters $\eta$ and
$\rho$ appearing in the 
Wolfenstein parametrization of KM mixing matrix. 

We include in this analysis the  interval  of values 
for $\Lambda_{\rm QCD}^{(4)}$ in \eq{lambdone}
and for the gluon condensate the range 
in \eq{Gscale}. The matching scale $\mu$ 
is varied between 0.8 and 1 GeV, while 
$\widehat B_K$ is varied according to 
the range obtained in the previous sections. 
The results for $m_t^{(pole)} = 180$ GeV are presented graphically in Fig. 7 .

\begin{figure}[t]
\epsfxsize=12cm
\centerline{\epsfbox{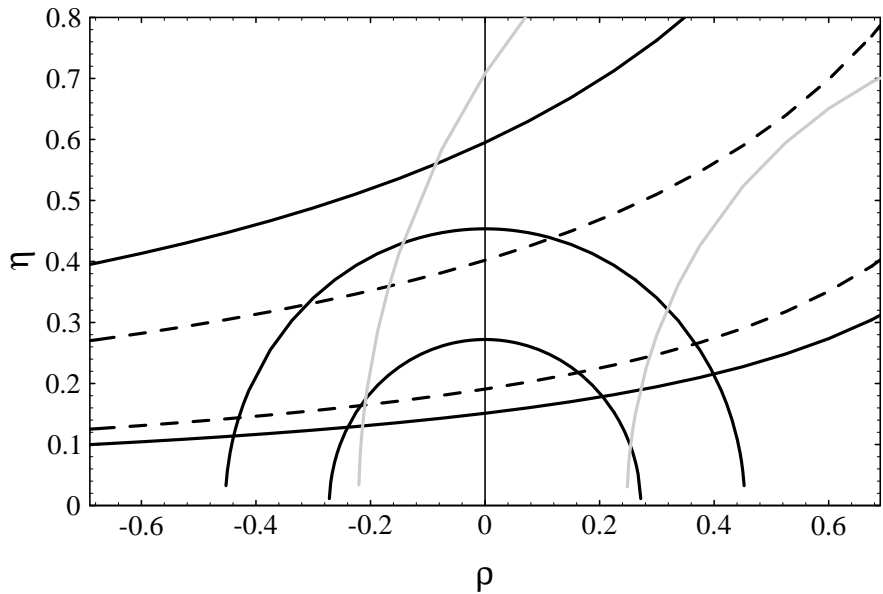}}
\caption{Constraints on KM parameters from kaon physics. See the text
for explanation.}
\end{figure}

We can see that the equations (\ref{bound1}) and (\ref{bound2}) define two 
families of curves (respectively hyperbola and circles) in the ($\rho$-$\eta$) 
plane. The allowed values of the two parameters correspond to the region 
delimited by the intersections between the two families of curves.
The area enclosed by the two solid line hyperbolae corresponds
to our most conservative range $0.54 < \widehat B_K < 1.2$. 

This procedure gives two possible ranges for $\eta$ and consequently 
for $\Im \lambda_t \simeq \eta |V_{us}| |V_{cb}|^2$, which correspond 
to having the KM phase in the I or II quadrant ($\rho$ positive or negative, 
respectively). 
For the central value of the top mass ($\overline{m}_t(m_W) \simeq 183$ 
GeV)  we find
\beq
 0.67\times 10^{-4} \leq \Im \lambda_t \leq 1.7 \times 10^{-4}
 \eeq 
 in the first quadrant and
 \beq
 0.41\times 10^{-4} \leq \Im \lambda_t \leq 1.7 \times 10^{-4}
 \eeq 
 in the second quadrant. 

We also consider a ``biased'' estimate of Im $\lambda_t$, 
obtained by fixing the gluon condensate and $\Lambda_{\rm QCD}^{(4)}$ to 
their central values and varying the matching scale $\mu$ between between 0.8 
and 1 GeV.  
In this way we are spanning the following range for the renormalization group 
invariant parameter
 $\hat B_K$ :
 \beq
 \hat B_K = 0.87 \pm 0.10 \, ,
 \label{BKrange} 
 \eeq

This choice of the input parameters 
restricts the hyperbolic band to the area enclosed by the dashed lines
in Fig. 7.
The overlap with the constraint of \eq{bound2}
leads to the following ranges of allowed values for Im $\lambda_t$
\beq
 0.81 \times 10^{-4} \leq \Im \lambda_t \leq 1.6 \times 10^{-4}
 \eeq 
 in the first quadrant and
 \beq
 0.52 \times 10^{-4} \leq \Im \lambda_t \leq 1.5 \times 10^{-4}
 \eeq 
 in the second quadrant.

Varying $m_t$ in the range $m_t^{(pole)} =180\pm 12$ GeV 
affects very little the quoted lower bounds on $\Im \lambda_t$
while the upper bounds are changed by less 
than 20\% (decreasing $m_t$ corresponds to increasing the upper limits).
On the other hand, the upper bound on $\Im \lambda_t$ remains stable, beeing 
bounded by the maximum value of $\eta$ obtained from \eq{bound2} ($\rho=0$).

We do not discuss here details of the bounds provided by $\Delta M_{B_d}$ 
which are represented, for the central value of $m_t$,
by the area delimited by the grey lines in Fig.~7.
The constraints of $B_d-\bar B_d$ mixing have a 
marginal impact in the determination of the overall range
of $\eta$. As the example in the figure shows, only the lower bound
of $\Im \lambda_t$ in the second quadrant is affected by such an inclusion
and raised to $0.6 \times 10^{-4}$.

\vspace{1cm}
\noindent
{\bf Acknowledgments}

This work was partially supported
by the EEC Human capital and mobility contract ERBCHRX CT 930132.

\appendix
\section{Feynman Rules for the $\Delta S=1$ Chiral Lagrangian}

We report the Feynman rules for the three terms (proportional to 
$G_{LL}^a$, $G_{LL}^b$ and $G_{\underline{8}}$) of the $\Delta S=1$ chiral 
lagrangian which are relevant for the calculation of the long-distance 
contribution $\Delta M_{LD}$. All momenta are entering the vertex.

\vspace{.5cm}\noindent
$G_{LL}^a$:
\bea
K^+ (p_1) \pi^- (p_2) 
&\quad &   \frac{2 i}{f^2} \ p_1 \cdot p_2 \  \nnu\\
K^0 (p_1) \pi^+ (p_2) \pi^- (p_3) 
&\quad &  -\frac{\sqrt{2}}{f^3} \ p_3\cdot (p_2 - p_1) \ \\
K^0 (p_1) K^+ (p_2) K^- (p_3) 
&\quad &  -\frac{\sqrt{2}}{f^3} \ p_2\cdot (p_1 - p_3) \ \nnu\\
K^0 (p_1) K^0 (p_2) \pi^+ (p_3) K^- (p_4)  
&\quad &  \frac{i}{f^4} \left[p_1\cdot p_2 
-\left(\frac{p_1 +p_2}{2}\right)\cdot\left(p_3+p_4\right)
+ p_3\cdot p_4 \right] \nnu
\eea

\vspace{.5cm}\noindent
$G_{LL}^b$:
\bea
K^0 (p_1) \pi^0 (p_2)   
&\quad &   \frac{i \sqrt{2}}{f^2} \ p_1 \cdot p_2 \  \nnu\\
K^0 (p_1) \eta (p_2)   
&\quad &   \frac{i}{f^2} \sqrt{\frac{2}{3}} \ p_1 \cdot p_2 \  \nnu\\
K^0 (p_1) \pi^0 (p_2) \pi^0 (p_3)  
&\quad &  - \frac{1}{\sqrt{2} f^3}  \left[p_1\cdot \left(\frac{p_2+p_3}{2}\right)
-p_2\cdot p_3\right] \  \nnu\\
K^0 (p_1) \eta (p_2) \eta (p_3)  
&\quad &  - \frac{1}{\sqrt{2} f^3}  \left[p_2\cdot p_3 -p_1\cdot \left(\frac{p_2+p_3}
{2}\right)\right] \ \nnu \\
K^0 (p_1) K^+ (p_2) K^- (p_3)  
&\quad &  - \frac{\sqrt{2}}{f^3}  (-p_1\cdot p_3 +p_1\cdot p_2)\  \\
K^0 (p_1) \pi^+ (p_2) \pi^- (p_3)  
&\quad &  - \frac{\sqrt{2}}{f^3}  (-p_1\cdot p_3 +p_1\cdot p_2)\  \nnu\\
K^0 (p_1) \eta (p_2) \pi^0 (p_3)  
&\quad &  - \frac{1}{\sqrt{6} f^3}  (p_1\cdot p_2 + 2 p_2\cdot p_3 -3 p_1\cdot p_3 )\ 
\nnu\\
K^0 (p_1) K^0 (p_2) \bar{K}^0 (p_3) \pi^0 (p_4) 
&\quad &   \frac{2 \sqrt{2}}{3} \frac{i}{f^4}  \left[p_3\cdot p_4 -
\left(\frac{p_1+p_2}{2}\right)\cdot p_4\right] \ \nnu\\
K^0 (p_1) K^0 (p_2) \pi^+ (p_3) K^- (p_4) 
&\quad &   \frac{i}{3 f^4} \left[4 p_1\cdot p_2 -2 \left(\frac{p_1+p_2}{2}\right)
\cdot\left(p_4+p_3\right)\right] \ \nnu\\
K^0 (p_1) K^0 (p_2) \bar{K}^0 (p_3) \eta (p_4) 
&\quad &   \frac{2}{3} \sqrt{\frac{2}{3}} \frac{i}{f^4}  \left[p_3\cdot p_4 -
\left(\frac{p_1+p_2}{2}\right)\cdot p_4\right] \ \nnu
\eea

\vspace{.5cm}\noindent
$G_{\underline{8}}$:
\bea
K^0 (p_1) \pi^0 (p_2)   
&\quad &   \frac{i \sqrt{2}}{f^2} \ p_1 \cdot p_2 \  \nnu\\
K^0 (p_1) \eta (p_2)   
&\quad &   \frac{i}{f^2} \sqrt{\frac{2}{3}} \ p_1 \cdot p_2 \  \nnu\\
K^+ (p_1) \pi^- (p_2)  
&\quad &  - \frac{2 i}{f^2} \ p_1 \cdot p_2 \  \nnu\\
K^0 (p_1) \pi^0 (p_2) \pi^0 (p_3)  
&\quad & - \frac{1}{2 \sqrt{2} f^3}  \left[p_3\cdot \left(p_1-p_2\right)+
p_2\cdot\left(p_1-p_3\right)\right] \  \nnu\\
K^0 (p_1) \pi^0 (p_2) \eta (p_3)  
&\quad &  - \frac{1}{\sqrt{6} f^3} (p_1\cdot p_3 +2 p_2\cdot p_3-3 p_1\cdot p_2) \ \nnu\\
K^0 (p_1) \eta (p_2) \eta (p_3)  
&\quad &  -\frac{1}{\sqrt{2} f^3} \left[p_2\cdot p_3 -p_1\cdot\left(\frac{p_2+p_3}{2}
\right)\right]\  \\
K^0 (p_1) K^+ (p_2) K^- (p_3)  
&\quad &  -\frac{\sqrt{2}}{f^3} \ p_3\cdot (p_2 - p_1) \ \nnu\\
K^0 (p_1) \pi^+ (p_2) \pi^- (p_3)  
&\quad &  - \frac{\sqrt{2}}{f^3}  \ p_2\cdot (p_1-p_3)\  \nnu\\
K^0 (p_1) K^0 (p_2) \bar{K}^0 (p_3) \pi^0 (p_4) 
&\quad &   \frac{2 \sqrt{2}}{3} \frac{i}{f^4}  \left[p_3\cdot p_4 -
\left(\frac{p_1+p_2}{2}\right)\cdot p_4\right] \ \nnu\\
K^0 (p_1) K^0 (p_2) \pi^+ (p_3) K^- (p_4)  
&\: &  \frac{i}{3 f^4} \left[p_1\cdot p_2 +\left(p_4 +p_3 \right)
\left(\frac{p_1+p_2}{2}\right)-3 p_3\cdot p_4 \right] \nnu\\
K^0 (p_1) K^0 (p_2) \bar{K}^0 (p_3) \eta (p_4) 
&\: &   \frac{2}{3} \sqrt{\frac{2}{3}} \frac{i}{f^4}  \left[p_3\cdot p_4 -
\left(\frac{p_1+p_2}{2}\right)\cdot p_4\right] \ \nnu
\eea
%
\clearpage
\renewcommand{\baselinestretch}{1}

\end{document}